\documentclass{desyproc}

\newcommand{\pom}{$\, \mathrm{I \!\!\! P} \, $}

\usepackage{subfigure}
\begin{document}
\title{Summary of the EDS Blois 2013 Workshop}

\author{{\slshape Michael Albrow$^1$}\\[1ex]
$^1$Fermilab. Wilson Road, Batavia, IL 60510, USA\\
Presented at EDS Blois 2013\\}



\acronym{EDS'09} 

\maketitle

\begin{abstract}
I give a personal overview of some highlights of the EDS Blois 2013 Conference on elastic and diffractive scattering in Saariselk\"{a}, Finland. 
\end{abstract}

\section{Introduction}
The true \emph{highlight} of the conference was, of course, the wonderful view of the aurora borealis on the night of the conference dinner!
But there were many other highlights in the 50 talks, so I must be very selective.
I apologize to the ultra-peripheral, heavy ion and cosmic ray speakers for not covering their fields. Names in [brackets] are citations to other
talks in the conference, in this volume, arXiv:1309.5705. The slides can be seen at http://www.hip.fi/EDS2013/

We are all familar with plots of the running of the strong coupling (do not say ``constant"), $\alpha_S$, vs. $Q^2$, and how $\alpha_S \sim$ 0.1 at 
$Q^2 = 10^4$ GeV$^2$ (where so many experiments ``test QCD"), but becomes about 0.5 at $Q^2 \sim$ 1 GeV$^2$. This is the realm of the \emph{real} strong
interaction, where perturbative calculations break down at large distances ($\sim$ 1 fm) and confinement sets in. We want to have a complete
understanding of the strong interaction among particles, hadrons as well as quarks and gluons. The oft-quoted remark that ``QCD is \emph{the} theory of
strong interactions" is justified at small distances, but I claim that if we cannot calculate a simple process like $pp$ elastic scattering, let alone
inelastic diffraction, we do not have a \emph{complete} theory. A consequence is that small-$Q^2$ physics is a region where new, even unexpected,
phenomena may be found. Approaches to understanding this field include lattice gauge theory, approximating the spacetime continuum with a lattice and
mainly calculating (with much effort) hadron spectra but not (yet) elastic scattering, string models of hadrons (out of fashion, but probably relevant),
and Regge theory, closer to our hearts. Regge theory has, I believe, not been taught in graduate schools for decades, but it is based on very sound
principles: Scattering amplitudes should be analytic (no sharps), obey unitarity (no probabilities $>$1) and crossing symmetry (what goes in can come
out!). For simple 2-body reactions such as $\pi^- + p \rightarrow \pi^0 + n$ it is the only game in town (exchange of the $\rho$-trajectory), and of course it
introduced us to the pomeron. Perhaps one day all the Regge phenomenology will be explained by QCD, or am I dreaming?
 Bjorken has said: ``Low $p_T$ is \emph{the} frontier of QCD". What happens to gluon fields at proton radii $>$ 1 fm? I asked him ``How can wee gluons, having
 large wavelengths, fit inside a Lorentz-contracted proton?" He said they don't, they are part of the vacuum and extend for $\sim$1 fm \emph{outside} the proton.
 Very interesting, because the pomeron, sometimes called the ``vacuum trajectory" has much to do with the (strong) vacuum. Double pomeron exchange
 can be called ``diffractive excitation of the vacuum". If we \emph{really} understood
 the vacuum we would understand all of fundamental physics; it is all in there if you probe deeply enough, and now LHC is probing the electroweak scale in the 
 vacuum (Higgs!). So exclusive $p+p \rightarrow p + H + p$ is diffractively exciting the vacuum (Higgs!) and allowing real H-quanta to pop out. 
 
 We now have new and precise measurements of the total $pp$ cross section $\sigma_T$, $d\sigma/dt|_{elastic}$, and from their difference $\sigma_{inelastic}$ 
 in $pp$ at the LHC by TOTEM, for $\sqrt{s}$ = 7 TeV and 8 TeV [Kaspar]. This allows a new \textsc{compete} fit to $pp$ and $p\bar{p}$ data from $\sqrt{s}$ = 10
 GeV to $10^5$ GeV (including some very approximate cosmic ray data).  At $\sqrt{s}$ = 10 GeV $\sigma_T(p\bar{p}) > \sigma_T(pp)$ and falling, 
 with $\sigma_T(pp) \sim$ 40 mb and flat. The rise of $\sigma_T$ was an early discovery at the first $pp$ collider, the CERN ISR, and is explained in
 Regge theory as due to the pomeron, \pom , trajectory having an intercept $\alpha_{I\!\!\!\!P}(t=0) >$ 1.0. The small and 
 decreasing difference $\sigma(p\bar{p}) - \sigma(pp)$ is
 due to the exchange of C = -1 reggeons, mainly $\rho_{R}$ and $\omega_R$ that have $\alpha_{I\!\!\!R}(t=0)\sim$0.5. At $\sqrt{s}$ = 10 (8000) GeV the rapidity gap between
 elastically scattered protons (= 2$\times y_{beam}$) is 4.73 (18.1). We see that t-channel exchanges over rapidity gaps $\Delta y \lesssim$ 5 still include
 reggeons, but  by $\Delta y$ = 8 (corresponding to $\sqrt{s}$ = 60 GeV) they are negligible, and the only exchanges left im elastic scattering at higher energy colliders are
 the \pom and the photon, which has $J = 1$ and does not disappear with increasing energy. Photon exchange (Coulomb scattering) is much smaller than
 \pom exchange because of the couplings, except at large impact parameters, $b \gtrsim$2 fm, corresponding to very small $|t|$, when strong
 exchanges(e.g. \pom ) have ranged out. The \textsc{compete} fit splits the difference between the Tevatron $\sigma_T$ measurements that disagreed by
 about 2$\sigma$, as could be expected.
 
 Now consider elastic scattering, which we all know is related to the total cross section by the optical theorem. If you like to think in pictures, my simple-minded, no equations
 explanation of the optical theorem is as follows. The total cross section is $pp \rightarrow X$ where $X$ is anything (summed over). Time reverse this
 and we have $X \rightarrow pp$, with the same amplitude (apart from complex conjugation). Stitch together and we have $pp \rightarrow X \rightarrow pp$,
 so inelastic scattering \emph{requires} elastic scattering and they are related. The stitched-together diagram resembles a ``ladder" exchange; if the poles and 
 rungs are quarks and gluons instead of hadrons it is a picture of the BFKL pomeron. Instead of all the hadrons in $X$ recombining to form $pp$ only
 some of them may, and we have a multiperipheral diagram of diffraction. 
 
 [Kaspar] showed the elastic scattering data of TOTEM now extending in to the Coulomb region $|t| < 10^{-3}$ GeV$^2$, and out to $|t| \sim$ 2 GeV$^2$. In
 the small-$|t|$ region Coulomb-nuclear interference tells us $\rho$, the ratio of real:imaginary parts of the forward scattering amplitude. Through dispersion 
 relations $\rho(s)$ tells us about the behaviour of $\sigma_T$ at much higher energies, constrained by the analyticity of
 scattering amplitudes $S(s,t)$.  For the first
 time we see an indication that $\rho$, which was rising (from negative values at low energies) and plateauing at $\rho \sim$  0.15 at Tevatron energies, may have 
 started to
 fall. Is this the first indication that $\rho \rightarrow 0$ at \emph{very} high energies, which would mean a purely imaginary scattering amplitude,
 saturation, the proton is a black disk [Dremin] and then $\sigma_{inel} = \sigma_{elastic} = \frac{1}{2} \sigma_T$? Let us hope for a confirmation 
 of a falling $\rho$ at
 $\sqrt{s}$ = 13 TeV! Moving to larger $|t|$, we see that the ``diffractive dip" that was discovered at the ISR around $t = -1$ GeV$^2$ has moved in to about
 -0.5 GeV$^2$ at 7 TeV; the effective proton radius (i.e. interaction range in $b$) has also increased, and the exponential slope beyond that dip has
 increased. The ISR, with its two independent rings, could also do $p\bar{p}$ collisions (as well as colliding deuteron- 
 and $\alpha$-beams)~\footnote{The ISR
 transformed our knowledge of hadron diffraction, with discoveries of the rising $\sigma_T$ which was the first evidence for the pomeron, high mass
 diffraction (above the resonance region), and double pomeron exchange, seen also in $\alpha\alpha$ collisions. As an aside, elastic $\alpha\alpha$ scattering at
 $\sqrt{s}$ = 126 GeV, shows a beautiful diffractive dip at $t$ = -0.10 GeV$^2$, discovered on-shift during data-taking!}. The dip is shallower 
 than in $p\bar{p}$ than in $pp$; evidence for some C-odd exchange.
 At still larger $|t|$, corresponding to very small transverse distances, all three valence quarks must be scattered in the same direction, 
 by (3-)gluon
 exchange. However prior to the new LHC data there were many predictions, with a factor $\sim$10 spread, none of which get full marks. [Islam] discussed a
 ``condensate-enclosed chiral-bag model", with the three valence quarks in a small ($\sim$0.2 fm) core, a shell of baryonic charge and an outer $q\bar{q}$
 condensate. He predicts a very flat $d\sigma/dt$ at large $|t|$ at 14 TeV. Can the small-core be tested another way? Double-parton scattering cross sections
 depend on transverse sizes, but can one select valence-quarks? How about ``double Drell-Yan" in $p\bar{p}$ Tevatron data? [Kohara] discussed elastic scattering amplitudes in ($t,b$)-space, predicting a dip at $t \sim$-4
 GeV$^2$ in $p\bar{p}$ at 1.8 TeV; unfortunately there is no data there. [Dremin] explained that a black disk model (which predicts, e.g., $\sigma_{elastic} =
 \sigma_{inelastic} = \frac{1}{2} \sigma_{T}$) is far away (if it is ever true), and there can be several gray-disk models with different evolutions of the proton's
 shape. The parton density at the proton's periphery increases with energy, and geometrical scaling is not valid.
 
 Beyond elastic scattering we have single diffractive excitation, SDE, with $p \rightarrow p^* \rightarrow X$. At low $M(X)$ it can be a resonance such as $N^*(1440)
 \rightarrow p\pi^+\pi^-$, which Good and Walker explained can be considered a component of the proton's wave function (or a momentary fluctuation $p \rightarrow
 p\pi\pi \rightarrow p$) made real by passage through the target (``selective filtering"). [Jenkovsky] discussed this; he expects a turn-down of 
 $d\sigma/dt dM$ at very small $|t|$, and emphasized that $d\sigma/dM$ (integrated over $t$) gets big contributions from baryon resonances, a warning to be very
 careful if trying to integrate diffrractive cross sections to low masses without data. It was discovered at the ISR that SDE extends to much higher masses than
 the resonance region, showing scaling behaviour in $M/\sqrt{s}$. Forward proton spectra show a diffractive peak for $x_{Feynman} >$ 0.95, approximately scaling,
 implying diffractive masses $M(X) \lesssim \sqrt{0.05} \sqrt{s}$ from 14 GeV at the ISR to 1800 GeV (!) at the LHC(8 TeV). Warnings: the inclusive proton spectra have not
 yet been measured at the LHC over this interesting region $0.90 < x_F < 1.0$, and the rapidity gap adjacent to an $x_F = 0.95$ proton is only $\Delta y$ = 3, where detectors
 are lacking. CMS has Forward Shower Counters, FSC, for $6 \lesssim |\eta| \lesssim 8$ which can be used in events with single interactions (no pile-up); they are
 simple scintillation counters that have so far been used for rapidity-gap tagging. They are being improved for Run 2 and should be useful for SDE
 measurements. At the Tevatron, CDF had Roman pots on the antiproton side; [Goulianos] showed the $t$-distribution of SDE $\bar{p}$ with a
 surprisingly flat $d\sigma/dt$ for $|t|$ = 2 - 4 GeV$^2$, and much higher than expected by the Donnachie-Landshoff model. Selecting the
 subset of events with two jets with $E_T >$ 20 GeV ($Q^2 \sim$ 900 GeV$^2$) one finds the same shape. CDF looked for ``exclusive dijets" by requiring a
 pseudo-rapidity ($\eta$) gap in the proton direction (with a $\bar{p}$ in the Roman pot) and calculating the ratio $R = M(JJ)/M(X)$ where $X$ is
 everything detected in CDF (excluding the $\bar{p}$). They find an excess over \textsc{pythia} without exclusive production, in fair agreement
 with the \textsc{exhume} prediction.

Diffraction was a major topic at HERA; initially something of a surprise (although predicted by Donnachie and Landshoff)
it became a key tool in understanding both protons and pomerons with a wealth of data. Soft (hadronic) diffraction can be seen as a radiated photon
(virtual or quasi-real) fluctuating to a vector meson, which scatters by pomeron exchange as in a real hadron-hadron collision.
Moving to higher $Q^2$ (perturbative regime), we have $\gamma^* \rightarrow q\bar{q} \rightarrow \gamma^*$, with the $q\bar{q}$ (a colour dipole)
scattering by 2-gluon-, a gluon ladder-, or BFKL pomeron-exchange. Measurements of diffractive structure 
functions [Valkarova] were done using forward proton spectrometers (FPS) or large rapidity gaps (LRG), with a 20\% difference
attributed to low mass, $M(X) <$ 1.6 GeV, diffractive excitation of the proton. The pomeron trajectory could be extracted by a Regge
fit, and the intercept $\alpha_{I\!\!\!\!P}(t=0) = 1.113 \pm 0.002^{+0.029}_{-0.015} $ is independent of $Q^2$ from 4 to 200 GeV$^2$.
[Kowalski] in his talk on low-$x$ physics stressed the correlations between the very low-$x$ region and the $x \rightarrow 1$
region, and its $Q^2$-dependence. There is new LHCb data [Malka] on elastic $J/\psi$ photoproduction, $\gamma + p \rightarrow J/\psi + p$,
extending the range of data from $W_{\gamma p}$ = 10 GeV to $>$1000 GeV. Extrapolation from HERA data works well, but the fixed target data
($W_{\gamma p} < 25$ GeV) have a steeper rise, indeed there seems to be a ``break" at about 25 GeV.

There is an active programme of diffractive studies at the LHC; I just select some examples. Both ATLAS [Monzani] and CMS [Goulianos]
have measured diffractive cross sections based on rapidity gaps within their detectors. It is important to specify exactly how one defines
``diffractive cross sections", as there is no absolute distinction between SDE and ND (= non-diffractive) events. If one had acceptance for the
\emph{complete} event (including leading protons) one might \emph{define} an event as diffractive if it has a proton with $x_F >$ 0.95 (but why not 0.96?)
or has a leading rapidity gap $\Delta y >$3 (but why not 4?). Neither ATLAS nor CMS can (yet) make such definitions, lacking the very forward
detectors. They can however measure the cross section for having a gap of length $\Delta\eta$ \emph{within} the detector, up to $\Delta\eta = 8$,
and they can do this for progressively tighter definitions of ``gap", which strictly speaking should be \emph{no hadrons} (photons are allowed, not
being strongly interacting). One should \emph{not} define a gap as having no jets above some $E_T$. Both CMS and ATLAS have shown that the cross section
as a function of gap width $\Delta\eta$ decreases until $\Delta\eta \sim$4, and then gently increases up to their limits $\Delta\eta \sim$8. This is expected
if the large gaps are dominated by pomeron exchange with the pomeron intercept $\alpha(t) >$1.0. A \textsc{pythia8} fit to the ATLAS data,
with a Donnachie-Landshoff flux parametrization, does not
describe the full distribution well, but the positive slope over $6 < \Delta\eta < 8$ gives $\alpha(0) = 1.058 \pm 0.003(stat) ^{+0.034}_{-0.039}$.
I find it remarkable that both elastic scattering and $X-GAP-Y$ events with high masses $X$ and $Y$ should be even close in \pom parameters.
When we have a two jets with a large rapidity separation (not gap) studies of azimuthal correlations ($\Delta\phi$), so-called Mueller-Navelet dijets,
teach us about BFKL dynamics, cascade models, etc. [Misiura]. Misiura also showed the remarkable jet $p_T$-spectra from 20 - 2000 GeV/c in rapidity bins, and
they agree with NLO predictions over 15 orders-of-magnitude! The high-$Q^2$ frontier of QCD is in good shape, so let us focus more on the low-$Q^2$ frontier!
Before I move there, Mesropian showed CMS results on diffractive (large rapidity gap) $W$ and $Z$ production, seen at the Tevatron but with much 
higher statistics at the LHC. Roughly 1.5\% of all $W$'s and $Z$'s have such big gaps, shown as the total energy$\sum E_T \lesssim$10 GeV in the forward calorimeters,
not well described by \textsc{pythia6,8} tunes. As $W$ and $Z$ come mainly from $q\bar{q}$ annihilation, this probes the quark content of the pomeron; an
alternative (but not incompatible) view would be that there is a $q\bar{q} \rightarrow W/Z$ annihilation followed (or preceded) by a colour exchange(s) such as 
to allow a big gap. 

  I now move to the low-$Q^2$ frontier. [Chung] discussed quarkonia, glueballs and hybrids, their quantum numbers and properties, and 
  stressed how much there is still to learn about hadrons (and thus, the strong interaction). [Astregesilo] and [Chung] showed very relevant COMPASS data, e.g. on
  $\pi^- p \rightarrow \pi^-\pi^+\pi^-p$, with an amazing $M(\pi^-\pi^+\pi^-)$ plot with up to 6$\times10^5$ events per 5 MeV bin, so smooth it looks like a 
  continuous line rather than a histogram. Partial-wave analyses confirm the assignments of many resonances in the COMPASS data.  I wish we had such high statistics in
  double pomeron exchange data (COMPASS is at too low $\sqrt{s}$ for that), which selects quantum numbers $I^GJ^{PC} = 0^+\mathrm{even}^{++}$! We could get there at the LHC with a dedicated (and optimized) week or two of
  low-pileup running. [Albrow] showed the CDF $\pi^+\pi^-$ spectrum between two $\Delta\eta >$ 4.6 gaps with 350,000 events, extending out to 5 GeV and with
  clear signals for $f_0(980)$ and $f_2(1270)$ with other possible structures, at both $\sqrt{s}$ = 900 and 1960 GeV. Other exclusive channels, such as
  $\eta(')\eta(')$ and $\rho\rho$ etc. are being analysed. [Harland-Lang] showed that the ``Durham Group" predicts that $\eta'\eta'$ 
  will be $\gtrsim 1000\times$
  higher than $\pi^0\pi^0$ at $M > 5$ GeV, because the $\eta,\eta'$ are flavour singlets and have a high glue content. This is an important measurement both
  as a test of the Durham model, but also to better understand ``gluey" mesons.
  LHCb is a very promising detector for this physics, especially with heavy flavours (exclusive $c\bar{c},b\bar{b}$); they are hoping to install forward shower
  counters (as in CMS) to improve the gap definition. They have already presented [Stevenson] the observation of (quasi-)exclusive $\chi_c \rightarrow J/\psi+\gamma$.
  LHCb have better mass resolution on $J/\psi+\gamma$ than CDF, who could not resolve the three $\chi_c$ states. They fit $M(J/\psi+\gamma$) to a sum of the three states and
  find that the $\chi_{c0}$ is only about 17\% of the observed events, which are dominated by $\chi_{c2}$. The latter has a (19.5$\pm$0.08)\% 
  branching fraction to $J/\psi +
  \gamma$, compared to only (1.16$\pm$0.08)\% for $\chi_{c0}$, nevertheless the ratio ($\chi_{c2}:\chi_{c0}$) is much higher than expected by Durham. We welcome STAR at 
  RHIC [Adamczyk] to the central exclusive field, showing their first $M(\pi^+\pi^-)$ spectrum at $\sqrt{s}$= 200 GeV. Common data with both TOTEM and CMS
  during low pile-up, high-$\beta^*$ running, has some (being analysed) low mass $h^+h^-$ data, but especially interesting is the sample of events with at least
  two jets, $E_T >$ 20 GeV, in CMS and one or both forward protons [Oljemark]. These dijet and trijet events are the cleanest ever seen at a hadron collider,
  and remind one of LEP events. But these dijets are nearly all $gg$, while at LEP there were all $q\bar{q}$.
  In a ``1-day, 100 bunches, $\langle N \rangle \sim$ 0.04" run, there are
  ``thousands" of $p + JJ$ (SDE) events and ``dozens" of $p + JJ(J) + p$ events. I am deliberately vague, the analysis is ongoing, but note that in
  ``10-days, 1000 bunches, $\langle N \rangle \sim$ 1.0" we could increase the statistics by a factor $>1000$. In $p + JJ(J) +p$ we need to measure the
  (small) exclusive fraction and its cross section, the trijet:dijet ratio, the $b\bar{b}$ fraction in the exclusive dijets, classify the trijets as
  $ggg$ or $q\bar{q}g$ (by topology) and in the $q\bar{q} g$ events measure the ratios $b\bar{b}:c\bar{c}:s\bar{s}:u\bar{u}:d\bar{d}$ (democracy?).
  This is quite a programme, but has good novel tests of QCD and of the backgrounds to be expected in later exclusive $H \rightarrow b\bar{b}$ searches.
  
    The idea of adding precision forward proton tracking and timing to search for exclusive H(125) production in high luminosity ($\langle N \rangle \sim 40$)
    running  is now at last moving ahead in CMS with TOTEM [Albrow,PPS]. The similar ATLAS Forward Proton (AFP) proposal [Sykora] will hopefully
    happen too. The CMS Collaboration Board have endorsed the physics program and detector concept, and a Technical
    Design Report, with TOTEM participants, is being  prepared. The first steps is to better integrate TOTEM and CMS triggers, readout and analysis so that any
    low-pileup running in 2015 (including $\beta^* = 90$m runs) can get more data than in 2012. At the same time new Roman pots will be installed to test
    their effect on high intensity beams: How close can they be to the beams, what backgrounds do they create, etc.? For 2016/2017 runs one hopes
    to have two arms with new high-luminosity instrumentation for $p+X+p$ physics up to 100 fb$^{-1}$. The Stage 1 detectors will be at $z \sim \pm 220$ m, not far enough for
    $p+H(125)+p$ (that is for Stage 2 (2018+?) at $z$ = 420 m) but opening up the exclusive (and not exclusive) jet physics to very high $E_T$ and, importantly, measuring
    $p+W^+W^-+p$. This process occurs via two-photon interactions and is very sensitive to quartic gauge boson couplings ($G_{\gamma\gamma WW}$)
    or $a^W_0,a^W_C$. Indeed CMS
    has recently published the observation of two candidate events ($e^{\pm} \mu^{\mp} E\!\!\!\!/_T$ with no other tracks on the $e\mu$ vertex), agreeing with
    the Standard Model and putting much more stringent limits than LEP did. Measuring both protons will allow this process to be studied at the highest
    luminosities (probably), also in the $e^+e^- E\!\!\!\!/_T$ and $\mu^+\mu^-E\!\!\!\!/_T$ channels, with good $M(WW)$ resolution and excluding the uncertainties of proton dissociation.
    Beyond Stage 1, stations at $\pm$420 m would allow the observation of $p+H(125)+p$, the ultimate in \emph{vacuum excitation}. After all the vacuum
    \emph{is} the Higgs field, so let us (diffractively) excite it. Before the H(125) was discovered, central exclusive production was promoted as a way to measure
    its mass, spin, CP and couplings to $b\bar{b}$, which are now becoming well known. Still, it is important to study this amazing particle every way we can, and
    even seeing it in CEP shows (uniquely?) that it \emph{must} have positive parity, P = 1, important to test.
    
    An important activity is the LHC Forward Physics Working Group, described by [Kepka]. All LHC experiments are involved, with theorists and
    phenomenologists, and will produce a CERN Yellow Report on forward physics, and define a common strategy for optimal running conditions.
    A strong case can be made for dedicated low-pileup running, special optics or not, not just for a few days but perhaps for about two weeks per year.
    In addition with the new generation of forward proton spectrometers we can look forward to exciting 
    studies of exclusive $WW,JJ,JJJ$, into the TeV region, and perhaps observe a heavier Higgs or unexpected 
    states.
    
    I apologise for not covering all of the interesting results, and not giving proper citations.
    I thank the organisers, especially Tuula Maki, for inviting me to give this summary and their support. 
I acknowledge funding from the U.S. Dept. of Energy.

\end{document}